\documentclass[aps,prl,twocolumn,english,floatfix,superscriptaddress]{revtex4-2}
\usepackage{amsmath, amsthm, amsfonts, amssymb,amstext}
\usepackage{mathptmx}
\usepackage{bm}
\usepackage{xfrac}
\usepackage{mathrsfs}
\usepackage{latexsym}
\usepackage{array}
\usepackage{braket}
\usepackage{amsmath,amssymb,amsfonts}
\usepackage{xcolor}
\usepackage{tikz,tkz-euclide}
\usepackage{dsfont}
\usepackage[colorlinks=true,citecolor=blue,linkcolor=blue,urlcolor=blue]{hyperref}

\def\beq{\begin{equation}}
\def\eeq{\end{equation}}
\def\barray{\begin{eqnarray}}
\def\earray{\end{eqnarray}}

\def\dd{\mathord{\rm d}}

\def\ee{\mathord{\rm e}}

\def\ii{\mathord{\rm i}}

\def\tr{\mathop{\rm Tr}}
\def\half{\textstyle\frac{1}{2}}

\def\k{\textrm{k}}
\def\a{\mathsf{ a}}
\def\f{{\rm f}}

\usepackage{graphicx}
\usepackage{dcolumn}
\usepackage{bm}

\begin{document}

\title{A quasi-particle picture for entanglement cones and horizons in analogue cosmology}
\author{C. Fulgado-Claudio}
 \email{carlos.fulgado@estudiante.uam.es}
\author{A. Bermudez}
\affiliation{Instituto de F\'\i sica Teórica, UAM-CSIC, Universidad Autónoma de Madrid, Cantoblanco, 28049 Madrid, Spain.}

\begin{abstract}
Although  particle production in curved quantum field theories (cQFTs) is  key to our  understanding of  the early universe and black hole physics, its direct observation  requires  extreme conditions or   unrealistic sensitivities. 
Recent progress in quantum simulators  indicates that  analogues of cosmological particle production  can be  observed in table-top  experiments of cold atomic gases described by effective cQFTs. 
This promises a high degree of  tunability in the synthesised curved spacetimes  and, moreover, sets a clear roadmap to explore the interplay of  particle production with other non-perturbative effects genuine to interacting QFTs. We hereby focus on  the appearance of scalar and pseudo-scalar condensates  for self-interacting Dirac fermions, and study how dynamical mass generation and  spontaneous symmetry breaking  affects  real-time dynamics through the lens of entanglement. 
We use  the entanglement contour (EC) to analyse the spatio-temporal  structure of particle production, showing that a quasi-particle picture for the EC captures the cosmological horizon in accelerating spacetimes, while also being sensitive to the effect of different symmetry-breaking processes. In particular, we show that the combined breakdown of time-reversal symmetry due to the expanding spacetime, and  parity due to a pseudo-scalar condensate, manifests through the structure of the  light-cone-like propagation of entanglement. 
\end{abstract}
\maketitle

\textit{Introduction.--} Quantum simulators (QSs) promise to become a transformative  tool to explore the complex quantum many-body frontier common to  various  disciplines of contemporary science~\cite{feynman1982simulating,cirac2012goals, Georgescu_2014}.  In   theoretical physics,  QSs can sometimes realise analogues of phenomena that would otherwise be difficult to observe in their original realm, if not altogether impossible. These two aspects come together in the study of quantum field theories (QFTs) in  curved spacetimes \cite{Birrell, Parker,PhysRevLett.85.4643,PhysRevA.70.063615,Barcel__2005,Schutzhold:2007mx,Horstmann:2009yh} by means of  experimental quantum devices \cite{Lahav_2010,Weinfurtner_2011,2019NatPh..15..785H,Kolobov_2021,Viermann_2022,_van_ara_2024} that can be used to explore  foundational questions about gravity, or test our current cosmological models of the universe. These QSs foster a genuinely multidisciplinary approach that integrates concepts from atomic physics, quantum optics and quantum information theory, with those of high-energy physics and general relativity. A particularly relevant example is  that of quantum entanglement \cite{Laflorencie_2016, Pasquale_Calabrese_2004, Calabrese_2009}, which  provides  insights into the role of  quantum correlations in the complexity frontier of many-body systems, and plays an important role in questions as diverse as the nature of the QFT vacuum or the evaporation of black holes, and also possibly in the information paradox for the structure of spacetime at the  quantum level.

In this Letter, we focus on QSs of ultra-cold atoms in Raman optical lattices~\cite{book_soc, PhysRevLett.110.076401,PhysRevLett.112.086401,PhysRevLett.113.059901,Wu83,PhysRevLett.121.150401,Songeaao4748,https://doi.org/10.48550/arxiv.2109.08885} as analogues of   particle production of self-interacting Dirac fermions in an expanding  spacetime with a tunable scale factor. We   demonstrate how the entanglement contour~\cite{Chen_2014}, when combined with the quasi-particle picture of entanglement dynamics~\cite{Calabrese_2005,Fagotti_2008,Alba_2017,Alba_2018} and a non-perturbative variational ansatz based on fermionic Gaussian states~\cite{PhysRevA.79.012306,Kraus_2010},  can provide a clean spatio-temporal resolution for the real-time  dynamics. Looking at particle production through  the entanglement contour, we unveil  interesting phenomena like a cosmological particle horizon in the entanglement spreading, and the impact of symmetry-breaking fermion condensates during particle production  on the entanglement-cone structure, pointing at   future research opportunities in analogue gravity \cite{Jacquet_2020}.

\textit{The model.}  We consider a  Friedmann-Robertson-Walker  spacetime  
$\dd s^2=\dd t^2-\mathsf{ a}^2(t)\dd\Sigma^2$ with  a  scale factor $\mathsf{ a}(t)$ and $\dd\Sigma^2$  depends on the metric of the spatial slice.   
In a cosmological scenario \cite{dodelson:2003}, the   scale factor   connects   an inflationary epoch to our current dark-energy and dark-matter dominated universe in $D=3+1$ dimensions. In contrast, in an analogue setup, we are free from these constraints, and can: {\it (i)} connect two asymptotic Minkowski regions $\mathsf{ a}_0\to\sf{a}_\f$ by an exponential scale factor controlled  by the Hubble rate $\mathsf{H}$, such that the duration is  $\Delta t=\frac{1}{\mathsf{H}}\log\frac{\mathsf{a}_{\rm f}}{\mathsf{ a}_0}$; {\it (ii)} vary the Hubble rate at will, as it is not fixed by cosmological observations, interpolating  between  adiabatic $\sf{H}\rightarrow0$ and  sudden $\sf{H}\rightarrow\infty$ expansions; and {\it (iii)} explore reduced dimensionalities that may lead to simpler analogue experiments, such as $\dd\Sigma^2=\dd{\rm x}^2$ in $D=1+1$ dimensions. This flexibility extends to the dynamical quantum  fields evolving in this expanding spacetime, which, in the standard cosmological model of inflation,  is driven by a scalar field governed by certain self-interaction potentials. For the reasons exposed below, we shall be interested in self-interacting Dirac spinors ${\psi}(x)\to{\psi}_i(t)$ discretised on a co-moving chain  ${x}_i=ia$, where $a$ is the lattice spacing  and $i\in\mathbb{Z}_{N_S}$. By using  conformal time via $\eta=\int {\rm d}t/\mathsf{ a}(t)$, and rescaling the    fields as $\psi_i\rightarrow\sqrt{\sf{a}(\eta)}\psi_i$, there is no need to simulate an inflating lattice with increasing proper distance  $d_{ij}(t)=\mathsf{a}(t)|x_i-x_j|$, and the regularised action reduces to a   fermionic QFT with explicit dependence of the microscopic parameters on conformal time. Finally, we can also {\it (iv)}  explore the subsequent evolution of the produced particles within a flat spacetime $\mathsf{a}_{\rm f}$, which will allow us to derive a neat picture of thermalisation as these particles propagate and fill in the universe.
 
 We work with a Hamiltonian lattice field theory that can be derived from  the original cQFT action 
 \beq
 S=\!\int\!{\rm d}^Dx\sqrt{-g}\,\overline{\psi}(x)\Big(g_{\mu\nu}(x)\tilde{ \gamma}^\mu\!(x)\nabla_\mu - m\!\Big)\! \psi(x)-V_{\rm int}(\overline{\psi},{\psi}),
 \eeq
 which includes a generic metric $g_{\mu\nu}(x)$ with a determinant $g={\rm det}(g_{\mu\nu})$ setting the volume element,  curved gamma matrices fulfilling $\{\tilde{ \gamma}^\mu\!(x),\tilde{ \gamma}^\nu\!(x)\}=2g^{\mu\nu}(x)$, and a covariant derivative $\nabla_\mu$ that contains the spin connection,  and the Christoffel symbols~\cite{FulgadoClaudio2023fermionproduction}. We also include fermion self-interactions through a potential $V_{\rm int}$ that depends on the fermion field and its adjoint $\overline{\psi}(x)=\psi^\dagger(x)\gamma^0$.  Working in conformal time and with the above rescaled spinors, the Hamiltonian adopts a very transparent form which can then be regularised in the above co-moving lattice. In particular, we  employ  a Wilson-type  regularisation to deal with fermion doubling \cite{PhysRevD.10.2445} which, in momentum space, yields a  free part  ${H}=\sum_{\textrm{k}\in\rm BZ}\psi_{\textrm{k}}^\dagger h_{\textrm{k}} \big(m\mathsf{a}(\eta)\big)\psi_{\textrm{k}}$ with 
\beq
h_{\textrm{k}}\big(m\mathsf{a}(\eta)\big)=-\frac{\sin \textrm{k} a}{a}\gamma^0\gamma^1+\left(m\a(\eta)+\frac{1-\cos \textrm{k} a}{a}\right)\gamma^0,
\label{eq:SPHam}
\eeq
where 
the gamma  matrices are $\gamma^0=\sigma^z$, $\gamma^1=\ii\sigma^y$ and $\gamma^5=\gamma^0\gamma^1$. 
This free  field theory has a simple interpretation as a tight-binding  model on a synthetic ladder, and  can be realised with minor modifications of recent experiments with alkaline-earth atoms in  Raman lattices \cite{https://doi.org/10.48550/arxiv.2109.08885}. Here,  the ladder legs represent the spinor components, encoded in two hyperfine levels of ${}^{87}{\rm Sr}$, and an energy imbalance that accounts for the time-dependent mass    $m\a(\eta)$, controlled by the  Raman detuning.
 
Being quadratic,  Wick's theorem holds for this part,  and the real-time dynamics can be   fully characterised through the correlation matrix $\Gamma_{ij}(\eta)=\bra{\psi(\eta)}\psi^{\phantom{\dagger}}_i\!\!\psi_j^\dagger\ket{\psi(\eta)}$. Regarding the full many-body Hilbert space, the dynamics is constrained to  a reduced manifold  of so-called fermionic Gaussian states (fGS) $\{\ket{\psi(\Gamma)}:\,\Gamma\in\textrm{GL}_{4N_S\!}(\mathbb{C})\}$ during the expansion. This  can then be used as a  variational family  to  approximate     genuinely-interacting QFTs. In this limit, particle production~\cite{Birrell, Parker}  can be 
formalised  by  a Bogoliubov 
transformation~\cite{bogo1, bogo2} that acts by mapping $\ket{\psi(\Gamma_0)}\mapsto \ket{\psi(\Gamma_{\rm f})}$ producing particles. Taking advantage of the above points {\it (i)-(iv)}, we can actually go beyond this simple limit and explore non-perturbative effects by including self-interactions according to the Gross-Neveu model~\cite{PhysRevD.10.3235}, namely  
\beq
{{V}}_{\rm int}=\frac{g_0^2}{2}\sum_i\left(\psi_i^\dagger\gamma^0\psi^{\phantom{\dagger}}_i\right)^{\!\!2}\!\!.\label{eq:Hint}
 \eeq
 Here, $g_0^2$ is the dimensionless coupling strength which, in contrast to the bare mass,    does not  get renormalised by the  scale factor. In a previous work \cite{fulgadoclaudio2024interactingdiracfieldsexpanding}, we have indeed shown that the fGS  ansatz   can be used to approximate the vacuum $\ket{\psi(\Gamma_0)}$, and that it coincides exactly with  large-$N$ predictions \cite{BERMUDEZ2018149}. The fGS predictions account  for the  onset of scalar  and pseudo-scalar   fermion condensates, which are specific elements of the correlation matrix of fermion bilinears  
 \beq
 \label{eq:condensates}
 \Sigma(\eta)=\frac{ g_0^2}{2aN_S}\sum_{\textrm{k}}\langle \psi_{\textrm{k}}^\dagger\gamma^0\psi_{\textrm{k}}\rangle,\hspace{1.5ex} \Pi(\eta)=\frac{\ii g_0^2}{2aN_S}\sum_{\textrm{k}}\langle \psi_{\textrm{k}}^\dagger\gamma^1\psi_{\textrm{k}}\rangle.
 \eeq
 During the expansion, these condensates will   evolve non-perturbatively affecting the particle production and vice versa,  
 which we can approximate variationally via our fGS, leading to     a set of self-consistent  non-linear differential equations 
 \beq
 \ii\frac{\dd}{\dd\eta}\Gamma(\eta)=\left[\tilde{h}_{\textrm{k}}\big(m\mathsf{a}(\eta),\Gamma(\eta)\big),\;\Gamma(\eta)\right].
 \label{eq:RTE}
 \eeq
 Here,  $h_{\textrm{k}}\to \tilde{h}_{\textrm{k}}\big(m\mathsf{a}(\eta),\Gamma\big)={h}_{\textrm{k}}\big(m\mathsf{a}(\eta)+\Sigma(\Gamma)\big)-\ii\Pi(\Gamma)\gamma^1.$
We see how the  non-linearities arise through $\tilde{h}_{\rm k}$ via  the self-energy  expressed in terms of  the  condensates,  which follows  from  Wick's theorem on the fGS manifold  $\bra{\psi(\Gamma)}{{V}}_{\rm int}\ket{\psi(\Gamma)}$.

Solving these  equations  with the initial condition  $\ket{\psi(\Gamma_0)}$ one obtains a time-evolved  fGS  $\ket{\psi(\Gamma(\eta_{\rm f}))}$  with a non-vanishing density of particles with respect to the instantaneous groundstate $\ket{\psi(\Gamma_{\rm f})}$ at $(m\mathsf{ a}_{\rm f},g_0^2)$. 
Within our variational approximation, the interpretation in terms of a Bogoliubov transformation is still useful, 
but it now depends on the full non-perturbative time history $\{\Gamma\}_\eta=\{\Gamma(\eta):\,\eta\in[\eta_0,\eta_{\rm f}]\}$, 
allowing to express the evolved state as
\beq
\label{eq:squeezed_state}
|\psi(\Gamma(\eta_{\rm f}))\rangle = \bigotimes_{\textrm{k} \in \text{BZ}} \left( \alpha_{\textrm{k}}(\{\Gamma\}_\eta)\ - \beta_{\textrm{k}}(\{\Gamma\}_\eta)\ a_{\textrm{k}}^\dagger b_{-{\textrm{k}}}^\dagger \right)|0\rangle, 
\eeq
with  normalised  Bogoliubov parameters  $|\alpha_{\textrm{k}}|^2+|\beta_{\textrm{k}}|^2=1$. 
The density of  produced fermions after the expansion is $n_a=n_b=\sum_{ \textrm{k}\in{\rm BZ}}|\beta_{\textrm{k}}(\{\Gamma\}_\eta)|^2/{{aN_S\mathsf{a}_{\rm f}}}$, which is the result of a back-to-back production of particle-antiparticle  pairs. 
We note that previous works \cite{BALL2006550,PhysRevD.81.084018,Fuentes_2010, Mart_n_Mart_nez_2014} on  free  QFTs have exploited the formal analogy  of this  transformation  with two-mode squeezing  to characterise mode entanglement in a  cosmological scenario in momentum space, capturing in this way correlations between right- and left-moving excitations. 
This, however, does not provide any spatial resolution that links to the causal propagation of entanglement, or to the existence of 
cosmological horizons due to an accelerated expansion.  Moreover, to our knowledge, there has been no prior study of the effect of interactions and fermion condensates  on the entanglement during gravitational expansion. We  fill in both gaps in this work.

\textit{Entanglement and the quasi-particle picture.--} Let us start moving away from the aforementioned mode entanglement by focusing on the block  entanglement entropy (EE)  $S_A=-\tr(\rho_A\log\rho_A)$, which quantifies the amount of quantum correlations  across the $AB$ bipartition  of the system $\rho_A=\tr_{B}\ket{\psi}\!\!\bra{\psi}$, where $A$ would be a block of $\ell_A$ sites. We note that the EE has been studied in great detail under the so-called quasi-particle picture. In its original realm, the quasi-particle picture describes the evolution of entanglement entropy  as carried by  propagating quasi-particles produced in pairs after a quantum quench. It was first introduced as a qualitative explanation for homogeneous quantum quenches in quantum chains \cite{Calabrese_2005}, and then quantitatively demonstrated for the XY model \cite{Fagotti_2008} and other generic integrable models \cite{Alba_2017,Alba_2018}. It has also been studied for other entanglement measurements beyond entanglement entropy, such as the so-called entanglement links \cite{Santalla_2023}. This picture suggests that, after a quantum quench, the initial state acts as a reservoir of excited quasi-particles that are created in pairs and propagate in straight lines with opposite velocities given by $v_\k=\pm|\partial_\k\varepsilon_\k|$, where $\varepsilon_\k$ is an energy dispersion relation for the quasi-particle labelled by $\k$. From this perspective, when the  partition involves two spatial regions $A$ and $B$, entanglement growth is due to entangled pairs of propagating quasi-particles residing in different sides of the partition. In our context of conformal time, the quasi-particle picture of the EE dynamics is
\beq
S_A(\eta)=\eta\int_{v_\k>\frac{\ell_A}{2\eta}}\frac{\dd \k}{2\pi}\,2v_\k s(\k)+\ell_A\int_{v_\k<\frac{\ell_A}{2\eta}}\frac{\dd \k}{2\pi}\,s(\k),
\label{eq:SM_entanglement_qp}
\eeq
where the first term describes an initial linear increase of entanglement over conformal time, which accounts for a phase in which quasi-particles are entering in the $A$ block.  
The second term, on the other hand,  describes an equilibration to an extensive amount of entanglement proportional to $\ell_A$ that appears for sufficiently long times, such that the same amount of quasi-particles enter as those that exit the block. Note that, if there is a maximum propagation velocity, the limits of the integrals define an effective causal cone for the spread of entanglement in the system. In this formula, $s(\k)$ denotes the amount of entanglement carried by the pair of quasi-particles, and can be obtained by analysing the long-time behaviour of the quantum state, which is expected to approach a generalised Gibbs ensemble (GGE) \cite{Alba_2021} with a well known thermodynamic entropy. Alternatively, this quantity can be calculated exactly for some integrable systems  by knowing the specific form of an initial squeezed state \cite{Bertini_2018}. 

Let us now adapt this quasi-particle picture to our cosmological scenario, noting that the  Hubble rate $\mathsf{H}$ allows us to smoothly interpolate between the adiabatic and quench limits. We note that after a antiparticle-hole transformation \cite{Schwartz_2013},  quasi-particles correspond to pairs of fermions and antifermions created back-to-back with opposite quasi-momentum, as described by Eq.~\eqref{eq:squeezed_state}. For this case, closed formulas can be found for the mode entanglement between particles and antiparticles, which are derived by tracing out the sector of antiparticles from the density matrix. We have
\beq
\rho=\bigotimes_\k\rho_\k=\bigotimes_\k\bigg (|\alpha_\k|^2|0\rangle\langle0|+|\beta_\k|^2|1_\k,\bar{1}_{-\k}\rangle\langle1_\k,\bar{1}_{-\k}|\bigg),
\eeq
where we have defined $|1_\k,\bar{1}_{-\k}\rangle={a}^\dagger_\k{b}^\dagger_{-\k}|0\rangle$. Then, tracing out the antiparticles sector, we arrive at
\beq
\rho^{\rm part}_\k=|\alpha_\k|^2|0\rangle\langle0|+|\beta_\k|^2|1_\k\rangle\langle1_\k|.
\eeq
The entanglement entropy between particles and antiparticles in the mode $\k$ is therefore given by the von Neumann entropy of $\rho^{\rm part}_\k$, $S(\rho_\k^{\rm part})=-\tr(\rho_\k^{\rm part}\log\rho_\k^{\rm part})$, which in terms of the Bogoliubov coefficients is given by
\beq
S(\rho_\k^{\rm part})=-|\alpha_\k|^2\log|\alpha_\k|^2-|\beta_\k|^2\log|\beta_\k|^2.
\label{eq:SM_eq_entropy_part_antipart}
\eeq
The relation between the Bogoliubov parameters for fermions, $|\alpha_\k|^2+|\beta_\k|^2=1$, allows to construct this expression using only the spectrum of particle production, with entanglement entropy being upper-bounded by $S(\rho^{\rm part})\leq\log2$ for $|\beta_\k|^2=\half$. Using this expression, the contributions $s(\k)$ in Eq.~\eqref{eq:SM_entanglement_qp} are found to be $s(\k)=2S(\rho^{\rm part})$, where the factor of 2 takes into account the two possibilities of a particle propagating out of the partition while an antiparticle stays inside and vice versa. 

\begin{figure}
    \centering
    \includegraphics[width=\linewidth]{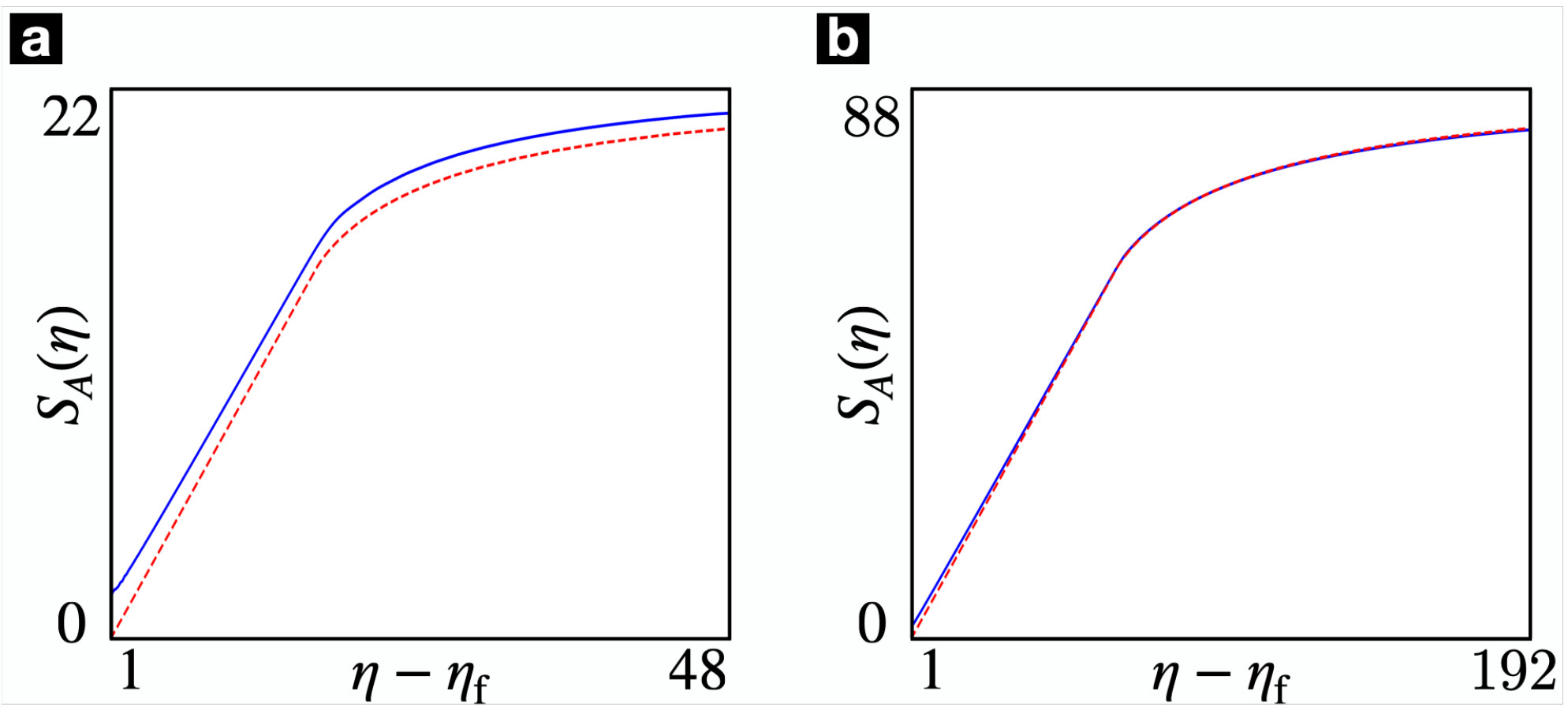}
    \caption{Time evolution of entanglement entropy $S_A(\eta)$ for a partition of {\bf(a)} $\ell_A=32$ sites from a $N_S=128$ sites system, and {\bf{(b)}} $\ell_A=128$ sites from a $N_S=512$ sites system. The blue line represents the numerical result while the red dashed line represents the quasi-particle prediction. The parameters are set as $ma=1$, $g_0^2=0$, $\a_0=0.01$ and $\a_{\rm f}=10$.}
    \label{fig:1}
\end{figure}

This quasi-particle  prediction \eqref{eq:SM_entanglement_qp} becomes exact in the non-interacting regime $g_0^2=0$ and in the scaling limit $\ell_A,\;t\to\infty$, while keeping $\ell_A/t$ constant. For large but finite $t$ and $\ell_A$,   it gives the leading order contribution for the EE dynamics of our propagating Dirac fields. In Fig.~\ref{fig:1} we benchmark the quasi-particle prediction with the exact time dynamics of the of the EE for two different values of $\ell_A$. We observe some deviations for small times, but a good agreement with Eq.~\eqref{eq:SM_entanglement_qp} for increasingly larger partition size as one approaches the scaling limit, according to the expected result. 

\textit{Entanglement contour and causal cones.--} One limitation of the block EE is the lack of spatial resolution even in the quasi-particle regime. This resolution  would be desirable as it would yield information about the real-time propagation of pairs of particle-antiparticle, giving insights  to the causal structure of the expanding spacetime. This limitation can be circumvented by resorting to the concept of entanglement contour (EC)~\cite{Chen_2014}, which assigns to each site $i$ of the partition a specific contribution $S_i$ to the overall EE, such that $\sum_iS_i=S_A$. With it, one can get further information on the local distribution of entanglement within the block, allowing us to bring this quasi-particle picture to a new domain. 
In particular, we  show below that the EC allows us to identify  a cosmological particle horizon due to the accelerating spacetime. 

Although a general recipe for the calculation of EC in arbitrary interacting models is lacking, 
the fGS ansatz provides a well-defined approximation to the EC 
via  the aforementioned correlation matrix $\Gamma(\eta)$, as 
the spectrum of the  block  density matrix 
$\rho_A$ is  determined by the restricted correlation matrix $(\Gamma_A(\eta))_{ij}=\Gamma_{ij}(\eta)|_{i,j\in A}$~\cite{Ingo_Peschel_2003, Surace_2022}. Diagonalising this matrix by a unitary transformation  $U^{\dagger}\Gamma_AU={\rm diag}(\{\nu_{\textrm{k}}\}_{\textrm{k}\in{\rm BZ}})$, one obtains an EC  $S_i(\eta)=\sum_{\textrm{k}\in{\rm BZ}}p_i(\textrm{\k},\eta)s_{\textrm{k}}(\eta)$ that weighs each contribution   $s_{\textrm{k}}(\eta)=-\nu_{\textrm{k}}(\eta)\log\nu_{\textrm{k}}(\eta)-(1-\nu_{\textrm{k}}(\eta))\log(1-\nu_{\textrm{k}}(\eta))$ at the $i$-th site  by $p_i(\k,\eta)=|U_{\textrm{k}i}(\eta)|^2$. Since this matrix is determined  self-consistently by solving for the fGS differential equations~\eqref{eq:RTE}, we can explore how   entanglement evolves in the spacetime  including interactions beyond perturbation theory, e.g.  symmetry-breaking fermion condensates. 

The quasi-particle picture can be extended to the entanglement contour, yielding the leading order contribution to the numerical results in the scaling limit. To do this, one starts by stating that pairs of particle-antiparticle are created homogeneously along the partition. Those created at sites closer to the edges will contribute earlier to the increase of entanglement, and effectively one sees entanglement stemming from both boundaries with a maximum velocity given by $v_{\rm max}=2\max_{\k\in{\rm BZ}}v_\k$. The factor of $2$ simply manifests the fact that both quasi-particles travel oppositely with the same velocity, so entanglement effectively travels at twice the speed of the individual quasi-particles.  This behaviour is captured by the following expression 
\beq
S_A(\mathbf{ x})=\int\frac{d\k}{2\pi}\,s(\k)\big(\Theta(2v_\k\eta-\mathbf{ x})+\Theta(2v_\k\eta-(\ell_A-\mathbf{ x}))\big),
\label{eq:SM_quasi-particle_EC}
\eeq
where $\Theta(x)$ is the Heaviside step function. This picture translates naturally to our lattice field theory as, by virtue of the particle-hole symmetry, EC is enforced to be invariant upon the exchange of the spinor components $\alpha\in\{u,\;d\}$, such that Eq.~\eqref{eq:SM_quasi-particle_EC} distributes evenly between them, $S_A({\bf x},\alpha)=\half S_A({\bf x})$.

\begin{figure}
    \centering
    \includegraphics[width=1\linewidth]{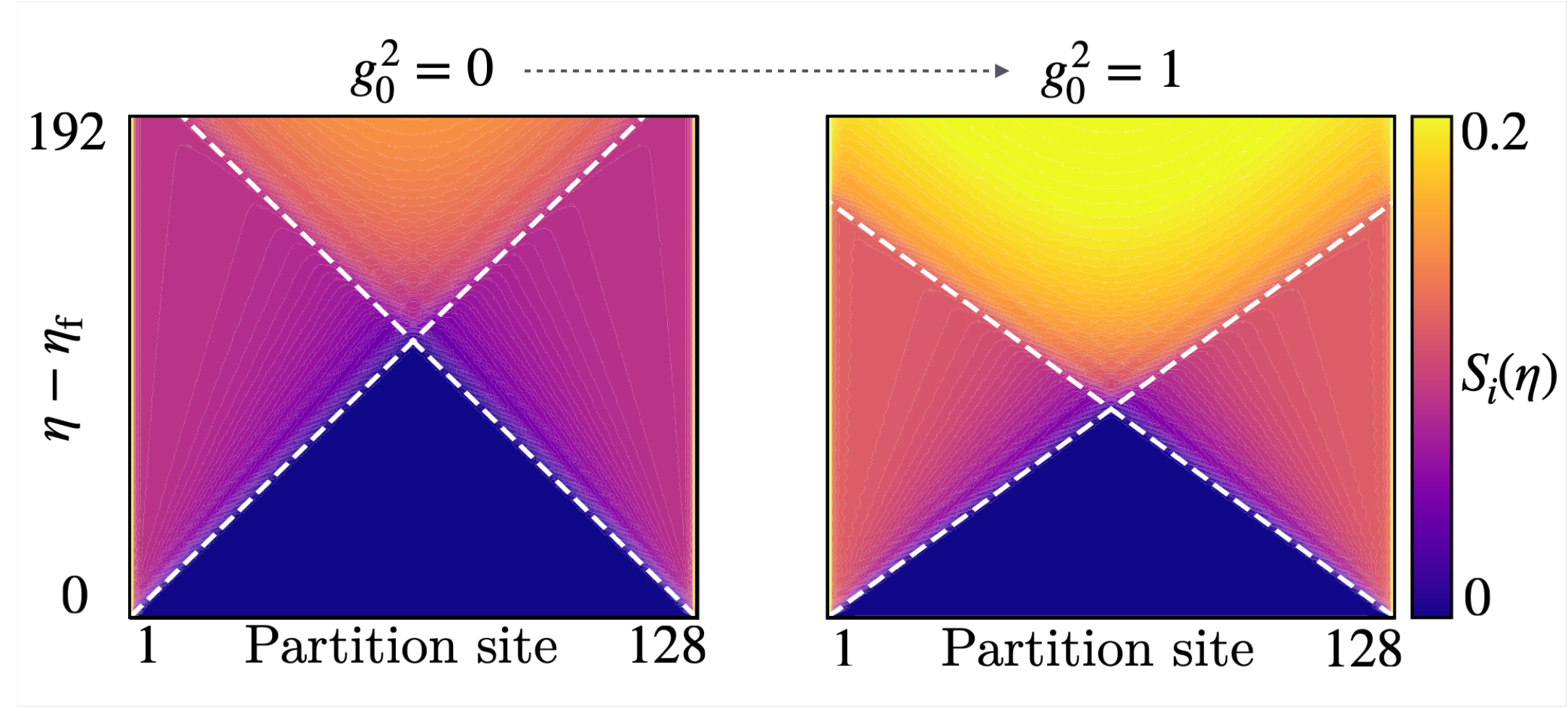}
    \caption{Spatio-temporal propagation of EC in the asymptotic flat region $\mathsf{a}_{\rm f}$, both in the free $g_0^2=0$ and in the interacting $g_0^2=1$ regimes, and setting $ma=-1$, $\a_0=0.7$, $\a_{\rm f}=1.3$, ${\sf H}a=100$ (sudden expansion). 
    The dashed white lines indicate the light-cone-like predictions based on the group velocity after the sudden expansion.}
    \label{fig:2}
\end{figure}

We start by studying how the fermion pairs  created during the expansion propagate in the asymptotically-flat  out region $\mathsf{a}_{\rm f}$, where the conformal time becomes linearly related to the cosmological time $\eta=\eta_\f+(t-t_\f)/\mathsf{a}_{\rm f}$ and can attain arbitrarily-large values. 
In Fig.~\ref{fig:2}, we depict the spatio-temporal  EC  profile  for a partition of $\ell_A=128$ sites  in a $N_{\rm S}=512$ lattice in real time. 
In the free case $g_0^2=0$, 
one observes a clear light-cone-like propagation of entanglement, which is consistent with the quasi-particle prediction in Eq.~\eqref{eq:SM_quasi-particle_EC}, which would be exact in the scaling limit, $\ell_A$, $t\to\infty$ with $t/\ell_A$ finite.
The cones emanate from the block boundaries, separating  causally connected and disconnected spacetime regions,  and show a slope that is set by twice the group velocity $v_{\rm g}=\max_{\textrm{k}\in{\rm BZ}}\left|\partial_{\textrm{k}} \epsilon_{\textrm{k}}\right|$, where $\epsilon_{\textrm{k}}$ is the energy of a particle or antiparticle obtained by diagonalising $h_{\rm k}$~\eqref{eq:SPHam}. Interestingly, this picture predicts that the entanglement equilibrates and there is an equipartition across the block $S_i(\eta)=S_0\,\,\forall i\in A$, such that the EE fulfils a volume law that agrees with the eigenstate thermalisation hypothesis~\cite{PhysRevA.43.2046,Srednicki_1994,Deutsch_2018} and a generalised Gibbs ensemble (GGE) that approximates $\rho_A$~\cite{Vidmar_2016}.

\begin{figure*}
    \centering
    \includegraphics[width=\linewidth]{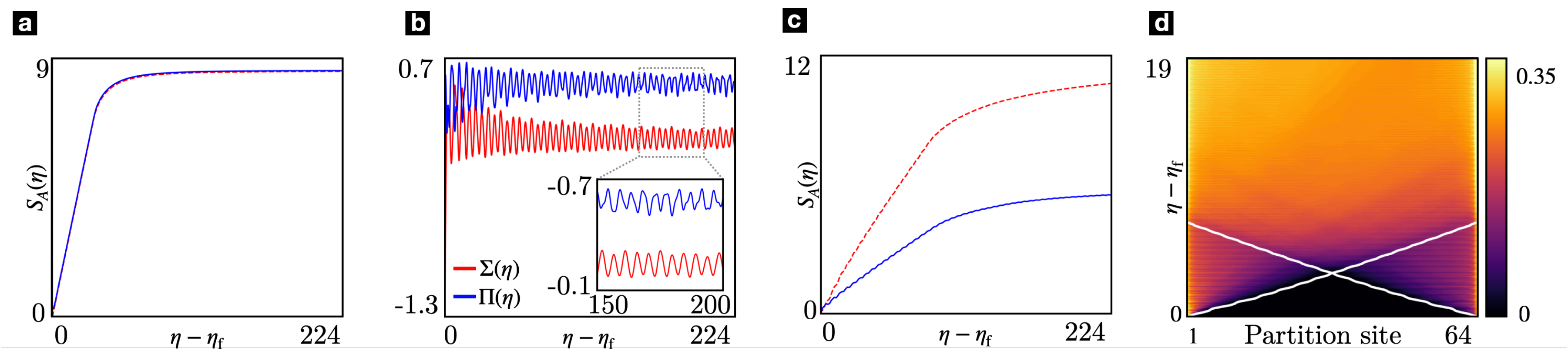}
    \caption{{\bf(a)} Evolution of entanglement entropy of a $\ell_A=64$ partition of a $N_S=512$ sites system after an expansion with a vanishing pseudo-scalar condensate $\Pi=0$. The blue line represents the numerical result while the red dashed line represents the quasi-particle prediction. The parameters are set as $ma=1$, $\a_0=0.01$, $\a_{\rm f}=10$, $g_0^2=3$ and ${\sf H}a=100$. {\bf(b)} Evolution of the scalar and pseudo-scalar condensates for a system of $N_S=512$ sites, showing their persistent oscillating behaviour. The parameters are set as $ma=-1$, $\a_0=0.7$, $\a_{\rm f}=1.3$, $g_0^2=3$ and ${\sf H}a=100$. {\bf(c)} Evolution of entanglement entropy of a $\ell_A=64$ partition of a $N_S=512$ sites system after an expansion with both a non-vanishing scalar and pseudo-scalar condensates, $\Sigma,\Pi\neq0$. The blue line represents the numerical result while the red dashed line represents the quasi-particle prediction.  The parameters are the same as in {\bf (b)}. {\bf(d)} Evolution of entanglement contour of a $\ell_A=64$ partition of a $N_S=512$ sites system after an expansion with both a non-vanishing scalar and pseudo-scalar condensates, $\Sigma,\Pi\neq0$. We show the contribution from the upper component of the spinor field to the EC, $S_i^u$. The white lines correspond to the curves $x(\eta)=\int^\eta v_g(\eta')d\eta'$. The parameters are the same as in {\bf (b)}.}
    \label{fig:3}
\end{figure*}

We now study the effect of a Gross-Neveu  self-interaction~\eqref{eq:Hint} on the EC, considering  first expansions with a vanishing pseudo-scalar condensate $\Pi(\eta)=0$~\eqref{eq:condensates}. In this case, the effect of interactions is to generate a scalar/chiral condensate that shifts the mass of the system dynamically $m\rightarrow m(\eta)= m+\Sigma(\eta)$. We find that, after a sudden period of exponential expansion $\mathsf{a}_0\to\mathsf{a}_{\rm f}$, the condensate develops some initial oscillations, after which it  equilibrates to a certain non-zero value. This results in a new dispersion relation from Eq.~\eqref{eq:RTE} that is  corrected by a time-dependent self-energy and can, consequently, lead to a dynamical group velocity $v_{\rm g}(\eta)$. In this case, the previous conserved quantities associated to the  GGE become time-dependent, and may compromise  the validity of the quasi-particle prediction. However, in the weakly-interacting regime, one expects to be able to  derive a hydrodynamic description for the EC  by  renormalising the spread velocity  and  the particle-antiparticle mode entanglement. We confirm this expectation in the right panel of Fig.~\ref{fig:2}, where we show that the EC  light cones follow  $x(\eta)\approx \pm2{v}^{\rm int}_g\eta$, where ${v^{\rm int}_g}$ is such a  renormalised velocity. We see that the corresponding causal structure in the EC displays a compressed cone and thus a faster propagation of quantum correlations. Since the initial bare mass is negative $ma=-1$, a positive and non-zero scalar condensate $\Sigma(\eta)>0$ actually makes the quasi-particles lighter, leading to faster propagation.  Although this is counter-intuitive at first, as the continuum QFT predicts a dynamical mass generation that would naively slow the dynamics, the problem on the lattice is richer as this dynamically-generated mass  actually brings one closer to a critical line around which the continuum limit is recovered. In this case, increasing interactions reduces the gap and also the renormalised bare mass, such that the particle-antiparticle pairs propagate faster. 

The initial oscillations can be  understood as the results of a finite-duration quench, after which the Hamiltonian renormalised by the self-energy equilibrates becoming static, and one can find constants of motion and recover the quasi-particle picture. The agreement with the quasi-particle picture in this interacting case  is shown in Fig.~\ref{fig:3}{\bf (a)}.
We note  that, due to the finite size of the system, some revivals are observed after the equilibration for $\eta\approx\frac{N_S}{2}$, and one has to numerically check the suitability of the quasi-particle picture by comparing the time scales of relaxation of $\Sigma(\eta)$, which is determined by the self-consistent time-evolution governed by Eq.~\eqref{eq:RTE}, and the time scales that one is able to explore numerically for a given system size.

We now move to  expansions with a non-vanishing pseudo-scalar condensate $\Pi(\eta)\neq0$. In this case, the scalar and pseudo-scalar condensates display some synchronised and persistent oscillations that do not equilibrate for the time scales that we have explored (see Fig.~\ref{fig:3}{\bf(b)}), and might connect to some long-wavelength hydrodynamic modes present at least in our fGS formalism. This leads to  long-lived oscillations of the number of particles and velocities of propagation. The lack of relaxation in this case restricts the applicability of the quasi-particle picture, as shown in Fig.~\ref{fig:3}{\bf(c)}, where we show that any attempt to fix $s(k)$ by their later time values leads to a completely incorrect prediction. With respect to the EC, the same limitations apply, with Eq.~\eqref{eq:SM_quasi-particle_EC} not being an accurate description of its numerical time-evolution. However, by virtue of a maximum velocity of propagation, one can still observe the presence of propagation cones in the evolution of EC, preserving the causal structure of the particle production phenomenon, as shown in Fig.~\ref{fig:3}{\bf (d)}. Additionally, in this case the spinor distribution of entanglement becomes non-trivial, yielding an interesting interplay with symmetry breaking as discussed in the next sections, and which is not predicted either by the quasi-particle picture. 

\begin{figure}
    \centering
    \includegraphics[width=1\linewidth]{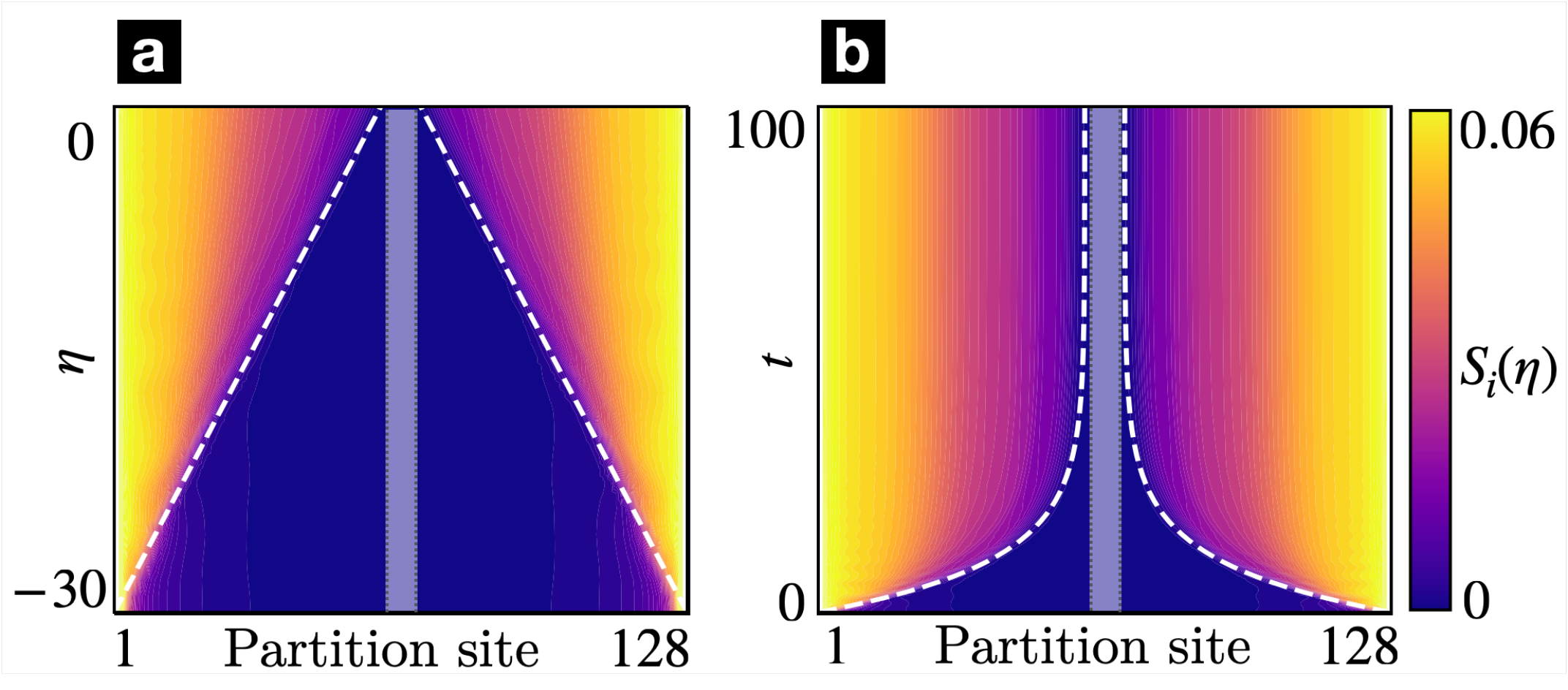}
    \caption{Spatio-temporal propagation of EC   during a de Sitter expansion, using conformal and cosmological time respectively, and setting $ma=1$, ${\sf H}a=0.1$, $g_0^2=2$, $\a_0=1/3$, with $\eta$ and $t$ expressed in lattice units. The expansion in  {\bf (a)} starts at $\eta_0a=-30$ and extends to the distant future  $\eta\to0^{-}$, approximated by $\eta a=-0.001362$. 
     The shaded regions denote the asymptotic  spatial boundary that particles cannot cross. 
     In  {\bf (b)}, the corresponding expansion  covers $t_0=0$ to  $ta\to\infty$, here approximated by 
    $t=100/a$.
    }
    \label{fig:4}
\end{figure}

{\it Entanglement  horizons.--}
Let us now turn to a key phenomenon  that becomes apparent via the EC dynamics 
during the accelerated de Sitter expansion $\mathsf{a}(\eta)$, and after it as above. For this cosmological expansion,  a particle horizon   determines  when the distance between the emission of a particle  and its subsequent detection  starts to expand faster than the speed of light.  Accordingly, spacetime breaks into causally disconnected regions that can have no influence on one another and, thus, share no entanglement. 
In conformal time, the appearance of this horizon in the EC is a bit subtle.  In Fig.~\ref{fig:4}{\bf (a)} we show the evolution of EC during the expansion, which 
shows a ballistic spread as before, 
but this  time is  upper bounded by $\eta=0$, 
as cosmological time  squeezes in the distant future  as a result of the exponential scale factor~\footnote{To ensure significant propagation since the beginning of the expansion, we impose an initial state with matter content. This is achieved via a quantum quench in the mass before the start of the expansion. Other alternatives to design the initial state would lead to similar interplay with the horizon.}. 

In Fig.~\ref{fig:4}{\bf (b)}, we show how resorting to  cosmological time $t$ allows for a transparent account of the above particle horizon. The EC profile in cosmological time  clearly shows how the propagation cones get curved as a consequence of the expansion of the universe, leading to a pair of asymptotes that 
 separate left-right regions that cannot be causally connected. We see two separated spatial regions where entanglement attains non-vanishing values as time elapses but, due to the accelerated expansion, these are separated by a finite region with a width $\Delta x=\ell_A-\frac{4v_g}{{\sf H}\a_0}$ in which the EC is always vanishing. We emphasise that the spatial resolution brought by the EC as opposed to the EE is crucial to identify these correlation horizons. It is also interesting to note that the EC distribution within the causally-connected parts does not fulfil an equipartition, so the evolution in this case is very different from an equilibration to a GGE. 
Had we let the system evolve in the Minkowski out region $\mathsf{a}_{\rm f}$, the particle-antiparticle pairs would  have had sufficient time to propagate in this static background, such that the causally-disconnected regions would again be connected by subsequent correlation cones as before, leading to volume-law entanglement and GGE.

\textit{ Symmetry breaking  in the entanglement contour.--} We have previously mentioned in passing that 
expansions 
in which the pseudo-scalar condensate is non-vanishing $\Pi(\eta)$ 
can  have a bigger impact on the spread of entanglement. We now   explore this further, 
considering that parity gets spontaneously broken, which will impact the  properties of the EC under point-like transformations \cite{Chen_2014}. Specifically, the EC satisfies that $S_i=S_j$ if the system is invariant under a transformation 
that 
exchanges two  sites, ${U}_{ij}\psi_i{U}_{ij}^{-1}=\ee^{\ii\xi}\psi_{j}$, where $\xi\in\mathbb{R}$  and $i\neq j$. 
To understand this property better, we temporarily  focus on  the discrete symmetries  of the static model $\a(\eta)=1$, which connects to the tenfold classification of topological insulators~\cite{Ryu_2010, Chiu_2016, Ludwig_2015}. For 
$\Pi=0$, the system is invariant under  time-reversal  ${T}$ and particle-hole  ${C}$  transformations, as well as their combination  ${S}={T}{C}$. At the single-particle level, these transformations are realised by  $\mathsf{T}=T\mathcal{K}$ with $T=\gamma^0$, $\mathsf{C}=C\mathcal{K}$ with $C=\gamma^0\gamma^1$, and $\mathsf{S}=\gamma^0\gamma^5$, where $\mathcal{K}$ denotes complex conjugation such that $T^\dagger h_{-\textrm{k}}^{*}T=h_{\textrm{k}}$, $C^\dagger h_{-\textrm{k}}^{*}C=-h_{\textrm{k}}$ and $\mathsf{S}^\dagger h_{\textrm{k}} \mathsf{S}=-h_{\textrm{k}}$. These operators satisfy $\mathsf{T^2=C^2=S^2}=1$. 
In addition, parity symmetry   ${P}$ is realised as  $P=\gamma^0$ such that $P^\dagger h_{-\textrm{k}} P=h_{\textrm{k}}$.  Indeed, together with translational invariance, parity symmetry describes the invariance under a mirror symmetry about any point and, in particular, about the center of the $A$ block,  which leads to a site exchange ${U}_{ij}$ with $j=\ell_A+1-i$. According to our previous comment, if parity is conserved, one expects the EC to display the symmetry $S_i=S_{\ell_A+1-i}$, which agrees with  the EC cones of Fig.~\ref{fig:2}.

\begin{figure}
    \centering
    \includegraphics[width=\linewidth]{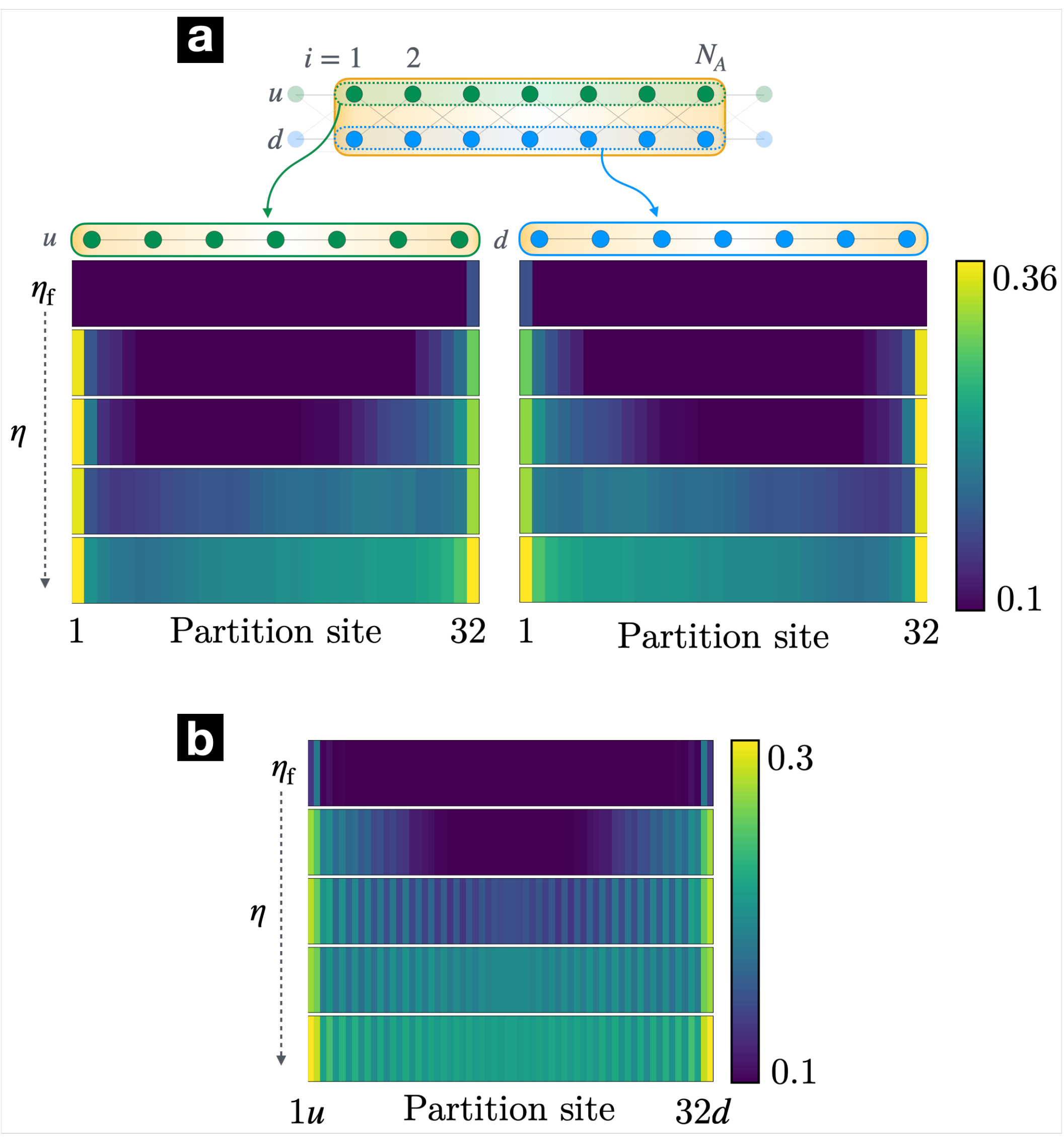}
    \caption{Time evolution of EC for different phases for a partition $A$ of $N_A=32$ sites, from a system of $N_S=128$ sites. {\bf (a)} Numerical time evolution of EC for each leg of the ladder, for an expansion with $\Pi\neq0$, with $ma=-1$, $g^2=0$, $\a_0=0.7$, $\a_\f=1.3$, $\mathsf{H}=100$ (quench limit). 
    {\bf (b)} Numerical time evolution of EC, with all the sites shown in the horizontal axis, alternating up and down spinors, for an expansion with $\Pi\neq0$, with $ma=-1$, $g^2=0$, $\a_0=0.7$, $\a_\f=1.3$, $\mathsf{H}=100$ (quench limit). 
    }
    \label{fig:5}
\end{figure}

This situation changes when $\Pi\neq0$, as one finds that ${T}$ is preserved, but ${P}$, ${C}$ and ${S}$  get broken. Moreover, switching back to $\mathsf{a}(\eta)\neq 1$  will also break ${T}$, endowing the condensate with its own dynamics $\Pi(\eta)$,  which shall affect the   EC spreading as we now discuss. In the left panel Fig.~\ref{fig:5}{\bf(a)}, we 
depict the spread of the EC in complete analogy to Fig.~\ref{fig:2}, but for an expansion with a non-zero $\Pi(\eta)$. 
We note that in all of the  cases above, the ECs displayed actually correspond to one of the spinor components, namely $S_{i}^{\rm u}(\eta)$, which does not compromise the generality as $S_{i}^{\rm d}(\eta)=S_{i}^{\rm u}(\eta)=: S_{i}(\eta)$.  This no longer holds when $\Pi(\eta)\neq0$, and one can observe in the right panel of Fig.~\ref{fig:5}{\bf(a)} that the EC for the down spinor   is indeed different  $S_{i}^{\rm d}(\eta)\neq S_{i}^{\rm u}(\eta)$, with a larger spread of entanglement from the left boundary with respect to the right one, opposite to what happens for $S_i^{\rm u}(\eta)$. These figures suggest that the EC profiles are still symmetric by  simultaneously exchanging left-right boundaries and up-down spinors,
yielding $S^{\textrm{u}}_i(\eta)=S^{\textrm{d}}_{\ell_A+1-i}(\eta)$ and vice versa $S^{\textrm{d}}_i(\eta)=S^{\textrm{u}}_{\ell_A+1-i}(\eta)$. 
In fact, one can check that this is indeed  a remnant symmetry that combines  both parity ${P}$ and particle-hole ${C}$ symmetries 
$(\gamma^1)^\dagger h_{\textrm{k}}^{*}\gamma^1=-h_{\textrm{k}}$, which still holds even if the individual symmetries are broken. The ${C}{P}$ symmetry that survives  a non-zero  $\Pi(\eta)$ suggests that one can define a zigzag ordering of the spatial sites and spinor components 
and depict the site- and spinor-resolved EC as shown in Fig.~\ref{fig:5}{\bf(b)}. This symmetry-aware representation allows to recover a clear light-cone-like structure characteristic of causality and connecting with the aforementioned quasi-particle picture, even in presence of a pseudo-scalar condensate and for any  Hubble rate $\mathsf{H}$.  

\begin{figure}
    \centering
    \includegraphics[width=0.8\linewidth]{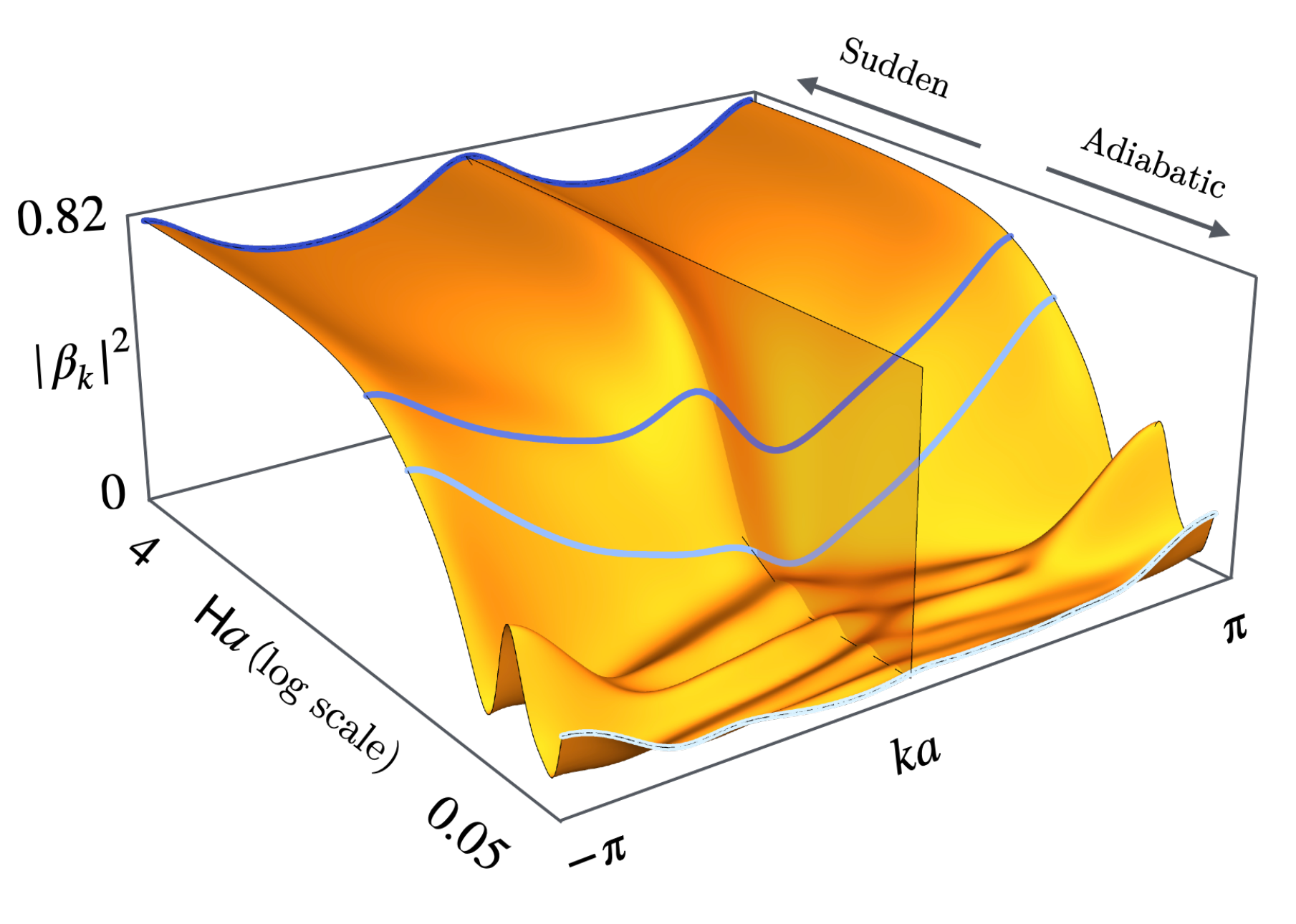}
    \caption{Spectra of particle production for various adiabaticity regimes. The parameters are set as $ma = -1$, $g_0^2 = 3$, $\a_0 = 0.7$, $\a_\f = 1.3$, for a chain of $N_S = 128$ sites. 
    We highlight four values of ${\sf H}a=4$, $0.3$, $0.2$ and $0.05$ for increasingly lighter blue, for which the relation between symmetry of the spectrum and $\mathsf{H}a$ is manifest.}
    \label{fig:6} 
\end{figure}

 Let us finish our discussion by showing how the resolution offered by the EC and its interplay with the broken symmetries contrasts the dependence on the Hubble rate $\mathsf{H}$ of other more standard quantities in cQFTs, such as the spectrum $|\beta_{\textrm{k}}|^2$ of produced particles $n_a\propto\int{\rm d}\textrm{k}|\beta_{\textrm{k}}|^2/2\pi$. In general,  the breakdown of parity  translates into  an asymmetric spectrum  $|\beta_{\textrm{k}}|^2\neq|\beta_{-\textrm{k}}|^2$, which is natural in light of its effect $\psi_{\textrm{k}}\rightarrow{P}\psi_{\textrm{k}}{P}^{-1}=\gamma^0\psi_{-\textrm{k}}$. However,  as shown   in Fig.~\ref{fig:6}, this is only manifested at intermediate values of the Hubble rate $\mathsf{H}$ in accordance with our previous results \cite{fulgadoclaudio2024interactingdiracfieldsexpanding}, but a symmetric spectrum around $\textrm{k}=0$ is recovered as one approaches the quench limit (see the darker blue curve). In the Supplemental Material \cite{supp}, we indeed show how this peculiarity can be understood from the interplay of parity with the above discrete symmetries, applied to the instantaneous state after the sudden quench.  
We unveil an additional underlying symmetry that protects an equal particle production in modes ${\rm k}$ and $-{\rm k}$ regardless of the breaking of parity that depends on the adiabaticity parameter $\mathsf{H}$, as it is only present in the quench limit and not for slower expansions.

\textit{Conclusions and outlook.--} 
Our study highlights the potential for observing exotic analogues of gravitational particle production in effective cQFTs that go beyond the standard cosmological model. The possibility of tuning the dimensionality, designing the expansion scale factor, and tailoring the matter content of the cQFT brings a wide and largely-uncharted territory that can be addressed from a multidisciplinary viewpoint. In this work, we have focused on self-interacting Dirac QFTs in a reduced Friedmann-Robertson-Walker spacetime, which have allowed us to explore how the typical phenomenon of particle-antiparticle production is affected by non-perturbative effects such as  dynamical mass generation and the appearance of symmetry-broken fermion condensates. Exploiting a multidisciplinary perspective, we have analysed this analogue cosmological setup through the lens of the quasi-particle picture of entanglement entropy and the  entanglement contour. Making use of non-perturbative variational techniques based on fermionic Gaussian states, we have  examined the spatio-temporal causal structure in the spread of entanglement carried by the produced particle-antiparticle pairs, and how the system can thermalise when looking at reduced parts. Our study reveals how the entanglement contour captures key phenomena, such as the cosmological particle horizon in a de Sitter spacetime,  and how the effects of symmetry-breaking processes become very transparent when resolving the symmetries of the entanglement contour. Our work paves the way for future studies exploring  further non-perturbative effects in interacting  cQFTs, and encourages  experimental efforts in Raman optical lattices to connect to these gravitational analogues.

{\it Acknowledgements.--}  We acknowledge support from PID2021- 127726NB-
I00 (MCIU/AEI/FEDER, UE), from the Grant IFT Centro
de Excelencia Severo Ochoa CEX2020-001007-S, funded
by MCIN/AEI/10.13039/501100011033, from the  the CAM/FEDER Project  TEC-2024/COM 84 QUITEMAD-CM, and from the CSIC Research Platform on Quantum Technologies PTI- 001.

\appendix

\section{Parity-breaking spectra and time-reversal symmetry}

Here, we show how the persistence of a symmetric spectrum of production in the quench limit for parity-breaking expansions follows from the persistence of time-reversal symmetry. As discussed in the main text, the Hamiltonian that governs time-evolution within our variational fermionic Gaussian state (fGS) ansatz is 
\beq
\begin{split}
\tilde{h}_{\textrm{k}}\big(m\mathsf{a}(\eta)\big)&=-\frac{\sin \textrm{k} a}{a}\gamma^0\gamma^1\\&+\left(m\a(\eta)+\Sigma+\frac{1-\cos \textrm{k} a}{a}\right)\gamma^0-\ii\Pi\gamma^1,
\end{split}
\label{eq:SM_SPHam}
\eeq
where $\Sigma={ g_0^2}\sum_{\textrm{k}}\langle \psi_{\textrm{k}}^\dagger\gamma^0\psi_{\textrm{k}}\rangle/{2aN_S}$, $\Pi={\ii g_0^2}\sum_{\textrm{k}}\langle \psi_{\textrm{k}}^\dagger\gamma^1\psi_{\textrm{k}}\rangle/{2aN_S}$ are respectively the scalar and pseudo-scalar fermion condensates. For values of the parameters for which $\Pi\neq0$, parity, particle-hole and sub-lattice symmetries are broken. Additionally, when considering an expanding universe with a time-dependent scale factor $\a(\eta)$, the remaining time-reversal symmetry must be analysed more carefully. First, one must select a specific value of time with respect to which the time reflection is performed. In our case, it is sensible to choose the initial time of the expansion $\eta_0$. 
Then, one finds that the condition for a Hamiltonian to be symmetric under time reversal becomes $T^\dagger h_{-\k}^{*}(\eta)T=h_{\textrm{k}}(2\eta_0-\eta)$. Clearly, the single-particle Hamiltonian in Eq.~\eqref{eq:SM_SPHam} with the scale factor under consideration does not respect this symmetry, as the expansion imposes a unique direction of time. However, the above condition is fulfilled instantaneously for $\eta=\eta_0$, so the properties that time-reversal imposes over the different observables are realised at $\eta=\eta_0$ and they disappear for $\eta>\eta_0$. 
Regarding the effect of these symmetries, while both parity and time-reversal transformations connect operators with opposite quasi-momenta, ${P}{\psi}_{\textrm{k}}{P}^{-1}={T}{\psi}_{\textrm{k}}{T}^{-1}=\gamma^0{\psi}_{-\k}$, only parity relates modes connected through a spatial reflection, ${P}{\psi}_{i}{P}^{-1}=\gamma^0{\psi}_{N_S+1-i}={T}\psi_{N_S+1-i}{T}^{-1}$. This manifests that ${P}$ is a unitary operator, while ${T}$ is anti-unitary. It is precisely these properties of time-reversal that underlies the results found in Figs.~\ref{fig:5}-\ref{fig:6} of the main text, which suggest the presence of a transformation that reflects quasi-momentum while leaving real space unaltered. Let us therefore explain the aforementioned results using the notion of time-reversal symmetry in this context. 

While parity is broken for expansions within the Aoki phase, time-reversal still plays a role at $\eta=\eta_0$. Since the instantaneous Hamiltonian ${H}(\eta_0)$ fulfils the time-reversal symmetry criterion, its groundstate $\ket{\rm gs_0}$ is time-reversal symmetric. As a consequence, when computing the density of produced particles, one finds that
\begin{equation}
\begin{split}
    |\beta_\k|^2&=\braket{{\rm gs}_0|{a}_\k^\dagger{a}_\k|{\rm gs_0}}\\&=\braket{{\rm gs}_0|{T}^\dagger{T}{a}_\k^\dagger{T}^{-1}{T}{a}_\k{T}^{-1}{T}|{\rm gs_0}}^{*}=\\&=\braket{{\rm gs_0}|{a}_{-\k}^\dagger{a}_{-\k}|{\rm gs}_0}=|\beta_{-\k}|^2,
    \end{split}
\end{equation}
even when parity no longer protects this symmetry. 
On the other hand, when the value of $\mathsf{H}a$ decreases and the expansion slows down, particle production is no longer computed over the initial groundstate $\ket{\rm gs_0}$, but over the evolved state given by Eq.~\eqref{eq:squeezed_state}. Since this state has evolved under the full time-dependent Hamiltonian which is consequently not time-reversal invariant, the above result is no longer fulfilled, leading to the asymmetric spectra in Fig.~\ref{fig:6} for small values of $\mathsf{H}a$. 

\bibliographystyle{apsrev4-1}
\bibliography{manuscript}

\clearpage
\onecolumngrid     
\section*{SUPLEMENTAL MATERIAL}

\section{Parity-breaking spectra and time-reversal symmetry}

Here, we show how the persistence of a symmetric spectrum of production in the quench limit for parity-breaking expansions follows from the persistence of time-reversal symmetry. As discussed in the main text, the Hamiltonian that governs time-evolution within our variational fermionic Gaussian state (fGS) ansatz is 
\beq
\begin{split}
\tilde{h}_{\textrm{k}}\big(m\mathsf{a}(\eta)\big)&=-\frac{\sin \textrm{k} a}{a}\gamma^0\gamma^1\\&+\left(m\a(\eta)+\Sigma+\frac{1-\cos \textrm{k} a}{a}\right)\gamma^0-\ii\Pi\gamma^1,
\end{split}
\label{eq:SM_SPHam}
\eeq
where $\Sigma={ g_0^2}\sum_{\textrm{k}}\langle \psi_{\textrm{k}}^\dagger\gamma^0\psi_{\textrm{k}}\rangle/{2aN_S}$, $\Pi={\ii g_0^2}\sum_{\textrm{k}}\langle \psi_{\textrm{k}}^\dagger\gamma^1\psi_{\textrm{k}}\rangle/{2aN_S}$ are respectively the scalar and pseudo-scalar fermion condensates. For values of the parameters for which $\Pi\neq0$, parity, particle-hole and sub-lattice symmetries are broken. Additionally, when considering an expanding universe with a time-dependent scale factor $\a(\eta)$, the remaining time-reversal symmetry must be analysed more carefully. First, one must select a specific value of time with respect to which the time reflection is performed. In our case, it is sensible to choose the initial time of the expansion $\eta_0$. 
Then, one finds that the condition for a Hamiltonian to be symmetric under time reversal becomes $T^\dagger h_{-\k}^{*}(\eta)T=h_{\textrm{k}}(2\eta_0-\eta)$. Clearly, the single-particle Hamiltonian in Eq.~\eqref{eq:SM_SPHam} with the scale factor under consideration does not respect this symmetry, as the expansion imposes a unique direction of time. However, the above condition is fulfilled instantaneously for $\eta=\eta_0$, so the properties that time-reversal imposes over the different observables are realised at $\eta=\eta_0$ and they disappear for $\eta>\eta_0$. 
Regarding the effect of these symmetries, while both parity and time-reversal transformations connect operators with opposite quasi-momenta, ${P}{\psi}_{\textrm{k}}{P}^{-1}={T}{\psi}_{\textrm{k}}{T}^{-1}=\gamma^0{\psi}_{-\k}$, only parity relates modes connected through a spatial reflection, ${P}{\psi}_{i}{P}^{-1}=\gamma^0{\psi}_{N_S+1-i}={T}\psi_{N_S+1-i}{T}^{-1}$. This manifests that ${P}$ is a unitary operator, while ${T}$ is anti-unitary. It is precisely these properties of time-reversal that underlies the results found in Figs.~\ref{fig:5}-\ref{fig:6} of the main text, which suggest the presence of a transformation that reflects quasi-momentum while leaving real space unaltered. Let us therefore explain the aforementioned results using the notion of time-reversal symmetry in this context. 

While parity is broken for expansions within the Aoki phase, time-reversal still plays a role at $\eta=\eta_0$. Since the instantaneous Hamiltonian ${H}(\eta_0)$ fulfils the time-reversal symmetry criterion, its groundstate $\ket{\rm gs_0}$ is time-reversal symmetric. As a consequence, when computing the density of produced particles, one finds that
\begin{equation}
\begin{split}
    |\beta_\k|^2&=\braket{{\rm gs}_0|{a}_\k^\dagger{a}_\k|{\rm gs_0}}\\&=\braket{{\rm gs}_0|{T}^\dagger{T}{a}_\k^\dagger{T}^{-1}{T}{a}_\k{T}^{-1}{T}|{\rm gs_0}}^{*}=\\&=\braket{{\rm gs_0}|{a}_{-\k}^\dagger{a}_{-\k}|{\rm gs}_0}=|\beta_{-\k}|^2,
    \end{split}
\end{equation}
even when parity no longer protects this symmetry. 
On the other hand, when the value of $\mathsf{H}a$ decreases and the expansion slows down, particle production is no longer computed over the initial groundstate $\ket{\rm gs_0}$, but over the evolved state given by Eq.~\eqref{eq:squeezed_state}. Since this state has evolved under the full time-dependent Hamiltonian which is consequently not time-reversal invariant, the above result is no longer fulfilled, leading to the asymmetric spectra in Fig.~\ref{fig:6} for small values of $\mathsf{H}a$. 

\end{document}